# Computation of three-dimensional displacements from the elastic strain field: application to mixed-mode indentation cleavage cracks in silicon

Abdalrhaman Koko[a1], T. James Marrow [a], and Elsiddig Elmukashfi[b]

[a] Department of Materials, University of Oxford, Oxford OX1 3PH, United Kingdom

[b] Department of Engineering Science, University of Oxford, Oxford OX1 3PJ, United Kingdom

## Abstract

A method has been derived to compute an experimentally measured strain field's equivalent three-dimensional displacement field by integration with Finite Element discretisation. The fields are approximated using piece-wise interpolation functions, with the deformation expressed as nodal displacements obtained using least squares optimisation and the assumption of linear elasticity.

The method is demonstrated in an analysis of $(\bar{1}10)$ and $(11\bar{1})$ cleavage cracks initiated at a micro-indentation impression on the (001) surface of a mono-silicon sample (single crystal) where the near-surface elastic deformation field was measured by high (angular) resolution electron backscatter diffraction. The displacement field defines the local boundary conditions of the residual field acting on cracks, and this has been used to calculate their three-dimensional stress intensity factors ($K_{I,II,III}$). The computed out-of-plane surface displacements are consistent with the topography measured by atomic force microscopy. However, the out-of-plane displacement depends on the (assumed) EBSD depth of information.

---

[1] Corresponding author. E-mail address: abdo.koko@materials.ox.ac.uk

# 1. Introduction

Fracture toughness at the microscale (e.g., of brittle coatings, particles and hard surfaces) is typically estimated by analysis of indentation-induced cracking via empirical equations, with assumptions of the crack geometry [1–3] and the loading history of the crack-tip during indentation [4]. For example, Lawn *et al*. [5–8] showed that the mode I fracture toughness ($K_{IC}$) can be estimated for a half-penny-shaped crack that is induced by Vickers indentation with knowledge of the contact impression radius ($a$), the surface crack length ($l$), Young's modulus ($E$), Hardness ($H$), and maximum indenter load ($P$), as in equation 1.

$$K_{IC} = x_v \left(\frac{E}{H}\right)^{1/2} \frac{P}{c^{3/2}}, \qquad c = l + a \qquad (1)$$

where $x_v$, is a fitting factor with a value of 0.015 ± 0.007 [2,9]. Dukino and Swain [10] modified equation (1) for Berkovich indentation. One potential drawback of using indentation to assess toughness is the uncertainty of the true subsurface crack shape, as the Palmqvist cracks that can initiate from the indentation impression corners [11]) appear similar to median or radial cracks [8]. The significant range of $x_v$ also leads to some inconsistency (and forced consistency [12]) in the toughness values obtained using indentation. Indentation cracking is complex, as can be illustrated by considering the case of silicon at room temperature. Silicon can deform in compression by a high-pressure phase transformation from a diamond cubic structure (Si-I phase) to a body-centred-tetragonal structure (β-Si or Si-II phase). The change in stress field with the application and removal of the indenter leads to elastic 'pop-in' and 'pop-out' by transformation-induced plasticity to the cubic BC8 (α-Si or Si-III phase) or rhombohedral R8 structure (Si-XII). Cracking occurs due to the resulting strain mismatch [13,14].

One in-depth review [15] concluded that – even for a highly finished surface – the indentation fracture method has "*poor value*" due to large scatter and user bias, and some believe that fracture toughness determined using indentation should not be accepted, and such data can only be used for comparative purposes [16]. Some of the issues – especially for anisotropic brittle materials – are dependency on the indenter tip angle [17–19], crack deflection [20,21], indentation depth/load dependency [22], and the effects of surface pile-up and size-dependent plastic deformation [23]. For further critical reviews on the indentation fracture method, the reader is referred to ref. [24–26]). Nonetheless, indentation methods have significant practical advantages since they enable materials testing at small length scales, so it is vital to improve confidence in their reliability. This may be aided by a better understanding the deformation fields that propagate cracks from indentations. Furthermore, the cracks that develop at the microscale within single grains on specific cleavage planes, or along grain boundaries, are not necessarily oriented perpendicular to the indented surface and may propagate under mixed-mode conditions. Such loading can be understood in terms of the displacements of the crack flanks that determine the three-dimensional stress intensity factors. Still, these displacements are quite difficult to measure in brittle elastic materials.

Our previous work on the full-field analysis of the deformation fields has shown that maps of the elastic strains local to the crack tip [27–30] can be analysed via a finite-element-based method to obtain their equivalent displacement fields. The finite element approach has also been applied by Friedman *et al.* [31] to compare the out-of-plane surface displacements and strain gradients around a wedge-indentation in mono-silicon, where these were derived independently using HR-EBSD and atomic force microscopy (AFM).

This analysis also quantifies the potential elastic strain energy release rate of quasi-static crack propagation via the *J*-integral, which can be decomposed to its equivalent stress intensity factors [32,33]. Hence the thermodynamic criteria for quasi-static crack propagation may be quantified by local measurements around the tip of a critical crack without knowledge of the external boundary conditions (i.e., applied load, crack length). The first such studies, with synchrotron X-ray diffraction [27–30], were done on mm-scale mode I cracks with low spatial resolution at a relatively large scale (cm-size specimens) via strain mapping in a two-dimensional (rectangular) field of view of regularly spaced square elements. Strain mapping to evaluate the J-integral and equivalent mixed mode stress intensity factors of stress-concentrating features can also be done at the microscale using high (angular) resolution electron backscatter diffraction (HR-EBSD), as demonstrated by both ex situ and in situ studies of slip bands [34–36], and twins [37,38], and cleavage cracks [31,39].

Hence, if measured using a suitably high-resolution method, the local displacement fields of a crack propagating from an indentation could be used to evaluate its critical strain energy release rate (and hence fracture toughness) without assumptions of crack length and geometry if the crack tip is in static equilibrium with the stress field. However, the stress field acting on a crack at an unloaded indentation is relaxed, particularly by lateral cracking [4,40,41], so its potential strain energy release rate is below the critical value for crack propagation [4]. Direct evaluation of the critical condition will require in situ observations under load. Still, ex situ studies of unloaded indentation cracks are worthwhile since establishing high confidence in their analysis is a necessary step toward more technically difficult in situ studies. In particular, a general method is required to evaluate cleavage cracks that are not necessarily orthogonal to the indented surface and are therefore subjected to mixed-mode loading (i.e., opening and shear). To address this issue, a robust method is required to derive the three-dimensional displacement field from a near-surface strain field measured at the micro-scale by high-resolution EBSD. The method must also have the flexibility to solve a non-uniform strain map with non-square elements, as this will facilitate efficient strain mapping experiments and also allow the direct use of the data as an input to standard Finite Element software codes (e.g., ABAQUS® [42]).

Here, a novel method for numerical integration of the elastic deformation field is derived and implemented. It uses the finite element method for discretising the field before assembling the boundary system of equations and solving for the nodal displacements using a least squares method. The method is scale independent, but as a demonstration, we calculate the 3D deformation field around cleavage cracks propagating from a Vickers indentation using high (angular) resolution electron backscatter diffraction (HR-EBSD). We then assess this by atomic force microscopy of the surface displacements. Finally, we use the displacement field calculated by the three-dimensional integration of the elastic strain field to obtain the residual stress intensity factors acting on cracks that the indentation has initiated.

## 2. Methodology

The numerical method for approximating the displacement from the measured strain or deformation gradient field is derived below (section 2.1). It is then applied to a measured HR-EBSD field around a Vickers indent impression in a mono-silicon crystal, which is discussed in the experimental part of the methodology (section 2.2).

## 2.1. Numerical method

Consider a class of problems in which the deformation field is given in a deformation measurement, such as the engineering strain or deformation gradient, in the deformed configuration. Hence, the scope of the proposed formulation is to determine the displacement field by integrating the deformation measure at a given set of points (i.e., mesh nodes). Thus, consider a body subjected to mechanical loading, which results in a deformation defined by the deformation gradient.

$$F_{ij} = \frac{\partial x_i}{\partial X_j} = \delta_{ij} + \frac{\partial u_i}{\partial X_j} \qquad 2$$

where $X_i, x_i$, 1,2,3, is a standard Cartesian coordinate system for the reference and deformed configurations, $u_i = x_i - X_i$ is the displacement vector, $\delta_{ij}$ is the second-order identity tensor and $H_{ij} = \partial u_i/\partial X_j$ is the displacement gradient. Thus, the displacement gradient can be split into infinitesimal strain $\varepsilon_{ij}$ (symmetric part) and rotations $\omega_{ij}$ (asymmetric part) that are given by:

$$\varepsilon_{ij} = \frac{1}{2}\left(\frac{\partial u_i}{\partial X_j} + \frac{\partial u_j}{\partial X_i}\right), \qquad \omega_{ij} = \frac{1}{2}\left(\frac{\partial u_i}{\partial X_j} - \frac{\partial u_j}{\partial X_i}\right) \qquad 3$$

The body can be discretised using finite elements to determine the nodal displacement field. Thus, the displacement field is interpolated as:

$$u_i(X_j, t) = \sum_{I=1}^{N_{\text{nodes}}} N_I(X_j) u_{iI}(t) \qquad 4$$

where $N_I$ are the standard finite element shape functions, which can also be expressed in terms of the parent element coordinates $\xi_i \in (\xi, \eta, \zeta)$, $N_{\text{nodes}}$ is the total number of nodes in the mesh and $u_{iI}$ are the values of the displacement fields. The reference and current configurations are respectively interpolated as:

$$X_i(X_j) = \sum_{I=1}^{N_{\text{nodes}}} N_I(X_j) X_{iI}, \qquad x_i(X_j, t) = \sum_{I=1}^{N_{\text{nodes}}} N_I(X_j) x_{iI}(t) \qquad 5$$

where $X_{iI}$ and $x_{iI}$ are the coordinates of node $I \in \Omega$ in the reference and current configurations, respectively. Using the definitions in equations (2) and 4), the displacement gradient can be obtained by:

$$H_{ij}(X_k,t)=\frac{\partial u_i(X_k,t)}{\partial X_j}=\sum_{I=1}^{N_{\text{nodes}}}\frac{\partial N_I(X_k)}{\partial X_j}u_{iI}(t)=\sum_{I=1}^{N_{\text{nodes}}}\frac{\partial N_I}{\partial \xi_l}\underbrace{\frac{\partial \xi_l}{\partial X_j}}_{J_{jl}^{-1}}u_{iI}(t) \tag{6}$$

where $J_{ij}=\partial X_i/\partial \xi_j$ is the mapping gradient from a reference configuration to the parent domain that can be determined from equation (7) as:

$$J_{ij}=\frac{\partial X_i}{\partial \xi_j}=\sum_{I=1}^{N_{\text{nodes}}}\frac{\partial N_I}{\partial \xi_j}X_{iI} \tag{7}$$

The mapping gradient of the current configuration to the parent domain is similarly determined from equation (8) as:

$$j_{ij}=\frac{\partial x_i}{\partial \xi_j}=\sum_{I=1}^{N_{\text{nodes}}}\frac{\partial N_I}{\partial \xi_j}x_{iI} \tag{8}$$

where $F_{ij}=j_{ik}J_{kj}^{-1}$. Hence, the displacement gradient can be written in terms of the current configuration as follows:

$$H_{ij}(X_k,t)=\sum_{I=1}^{N_{\text{nodes}}}\frac{\partial N_I}{\partial \xi_k}\frac{\partial \xi_l}{\partial x_j}F_{lk}u_{iI}(t) \tag{9}$$

The expressions (6) or (9) contain a system of equations in which the unknowns are the nodal displacements $u_{iI}$ that are of a number $N_{\text{Un}}=N_{\text{nodes}}\times N_{\text{Dim}}$, where $N_{\text{Dim}}$ is the number of dimensionality of the problem. The displacement gradient in the left-hand side can be prescribed anywhere within the element. Therefore, a set of computational points (i.e., $N_{\text{P}}$ points) that lie within the element can be chosen. Thus, the total number of equations becomes $N_{\text{Eq}}=N_{\text{Ele}}\times N_{\text{P}}\times N_{\text{d}}$, where $N_{\text{Ele}}$ is the total number of elements in the mesh and $N_{\text{d}}$ is the number of components of the deformation measure, e.g., in 3D problems, $N_{\text{d}}=9$. The system of equations can be written in the algebraic form in (10)[2].

---

[2] Equation (10) is the well-known $F=KU$ equation for the finite element method, where $F$ is the force vector, $K$ is the elements stiffness matrix and $U$ is the displacement vector. More details in supplementary information: **Error! Reference source not found.**.

$$A_{ij} d_j = b_i \quad (10)$$

where $A_{ij}$ is a $N_{\mathrm{Eq}}$ x $N_{\mathrm{Un}}$ coefficients matrix, $d_j$ is a vector that encapsulates the unknown nodal displacements $u_{iI}$, $b_i$ is a vector that encapsulates the displacement gradient $\partial u_i / \partial X_j$. The linear algebraic equations in (10) result in one of three solution sets:

(i) If $N_{\mathrm{Eq}} = N_{\mathrm{Un}}$ this will lead to a single unique solution. Thus, $A_{ij}$ has a full rank, and its inverse $A_{ij}^{-1}$ is unique.

(ii) If $N_{\mathrm{Eq}} > N_{\mathrm{Un}}$ this will make the system of equations overdetermined, and the solution can be best solved using the least-squares method. Thus, a minimisation of the square error can be written as below:

$$\min \left\| A_{ij} d_j - b_i \right\|^2 \quad (11)$$

The solution to the minimisation problem of the values of the nodal displacements $d_j$ is then obtained by equation (12).

$$d_j = \left( A_{ij}^T A_{ij} \right)^{-1} A_{ij}^T b_i \quad (12)$$

(iii) If $N_{\mathrm{Eq}} < N_{\mathrm{Un}}$ this will make the system of equations underdetermined with an infinite number of solutions unless the problem is subjected to a constraint or regularisations (e.g., condition matrix when using pseudoinverse method ($\mathrm{cond}(A_{ij})$) $= \left\| A_{ij} \right\| \left\| A_{ij}^{+} \right\|$, where $A_{ij}^{+}$ is the Moore–Penrose inverse of an $A_{ij}$ matrix).

This method's implementation and benchmarking using synthetic two- and three-dimensional data can be found in the supplementary information.

## 2.2. Experimental Method

A pre-polished single crystal Silicon sample was micro-indented on the surface (001) plane at room temperature using a Vickers Diamond Pyramid indenter (136° between faces), loaded with 50-gramme force (gf) for 1 second. These conditions are suitable for initiating half-penny cracks with minimal chipping or radial-cracks. The unloaded sample was then fixed to an aluminium stub using conductive paste (Silver DAG), which minimises image drift and the intensity of blooming caused by electron beam charging. The sample was subsequently attached to a Universal EBSD 70° pre-tilted sample holder and placed inside a Carl Zeiss Merlin field emission gun scanning electron microscope (FEG-SEM) for the acquisition of high-quality 800 x 600-pixel electron backscattering patterns (EBSPs) in a conventional EBSD setup (Figure 1a) using Bruker eFlash CCD camera. The operating parameters were 20 kV/10 nA beam condition, 18 mm working distance, 200 millisecond exposure time per pattern, 4 x 4 hardware pattern binning, and 0.25 µm step size.

The elastic deformation of the surface was calculated using a cross-correlation-based analysis (high angular resolution electron backscatter diffraction [43,44]) on the collected EBSP with a reference pattern chosen remotely from the stress concentration [45]. The independent change in interplanar spacing and shifts in zone axes were measured in 30 regions of interest (ROI) in each EBSP with a bicubic interpolation method for the best-fit solution. The measured small shifts and distortion between EBSPs and reference EPSP were then related to the elastic displacement gradient ($\nabla u^e$) and polarly decomposed to deviatoric strains (symmetric part, where $ij = ji$), $\varepsilon_{ij}$, and lattice rotations (asymmetric part, where $ii = jj = 0$), $\omega_{ij}$. Stress was calculated using (001) Silicon anisotropic elastic constants of $C_{11} = 165.7, C_{44} = 79.6, C_{12} = 63.9$ in GPa [46]. For the 18 dislocation types (12 edge and six screw dislocation systems) of face-centric cubic (FCC) mono-Si crystal, the geometrically necessary dislocation (GND) density was also estimated from the lattice rotation [47].

Hexahedron (brick) elements with eight nodes were then used to structurally mesh the field with the assumption of thickness Z of the membrane layer probed by the backscattered electrons. The model excluded the cracks (Figure 3a), which were identified by their locally elevated level of GND density (>1.6 x $10^{13}$ $m^{-2}$). The thickness Z was obtained by simulating the EBSP acquisition process using a Monte Carlo Simulation (Casino v2.48 [48]) of the trajectory of 5 million electrons impinging a bulk Si sample (Figure 1a) with a beam radius of 25 nm at these conditions [49]. The probabilities of the (EBSP) signal being from a certain depth was binned (Figure 1b) into ranges of 1, 2, 5, 8, 11, 16, 20, 40, 60, 80, 100, 130, 173, 210, 280, 350, 450, 570, 700, 850, 1100, and 1600 nm with the mean (50% probability) being at 173 nm and the mode at 40 nm. The strain field was then integrated into the equivalent elastic displacement field for these values of Z.

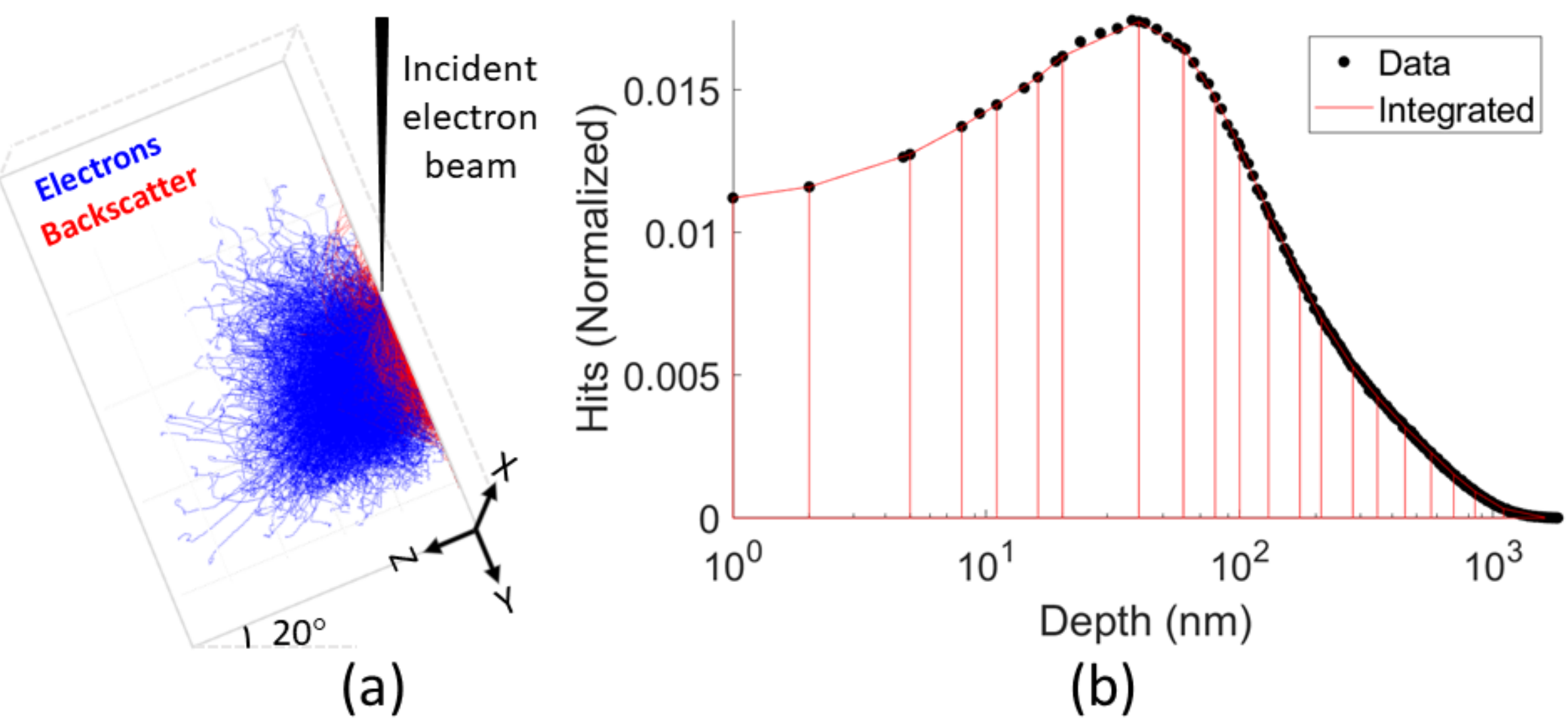

Figure 1: (a) Monte Carlo simulation of electrons trajectory. (b) Probability of backscattered electrons.

The topography around the indentation contact impression was measured using a Veeco AutoProbe (high-resolution) atomic force microscopy (AFM) in contact mode using a 10 nm probe tip and < 1 nm accuracy with a scan speed of 0.35 line/sec and 15.6 nm step size for an 8 x 8 µm field of view.

The subsurface geometry of the cracks was revealed using Focused Ion Beam (FIB) slicing. The indented sample was placed inside a Zeiss Auriga dual-beam SEM-FIB system with a Schottky field emission Gemini electron column and an Orsay Physics “Cobra” Ga+ ion FIB. The sample was tilted after achieving eccentricity at 54° before moving to a working distance of 5 mm. The FIB and electron beam coincidence was achieved by adjusting the stage-beam working distance and spatial stage movement, allowing simultaneous milling and secondary electron imaging (SEM). Once eccentricity and coincidence were achieved, a protective ~1.5 µm platinum and ~1.5 µm carbon layers were deposited using a 240 pA/ 30 kV beam to protect the surface, as shown in Figure 2.a. A 35 µm deep trapezium trench was milled using 16 nA/

30 kV to allow for easy viewing of the feature in the 3rd dimension (Figure 2b). In-lens and Secondary Electron (SE) Imaging conditions with 36° tilt correction (effective 90° viewing) were used for fine milling into the indentation contact impression (green arrow in Figure 2a) using ATLAS 3D with 600 pA/ 30kV milling conditions (Figure 2c).

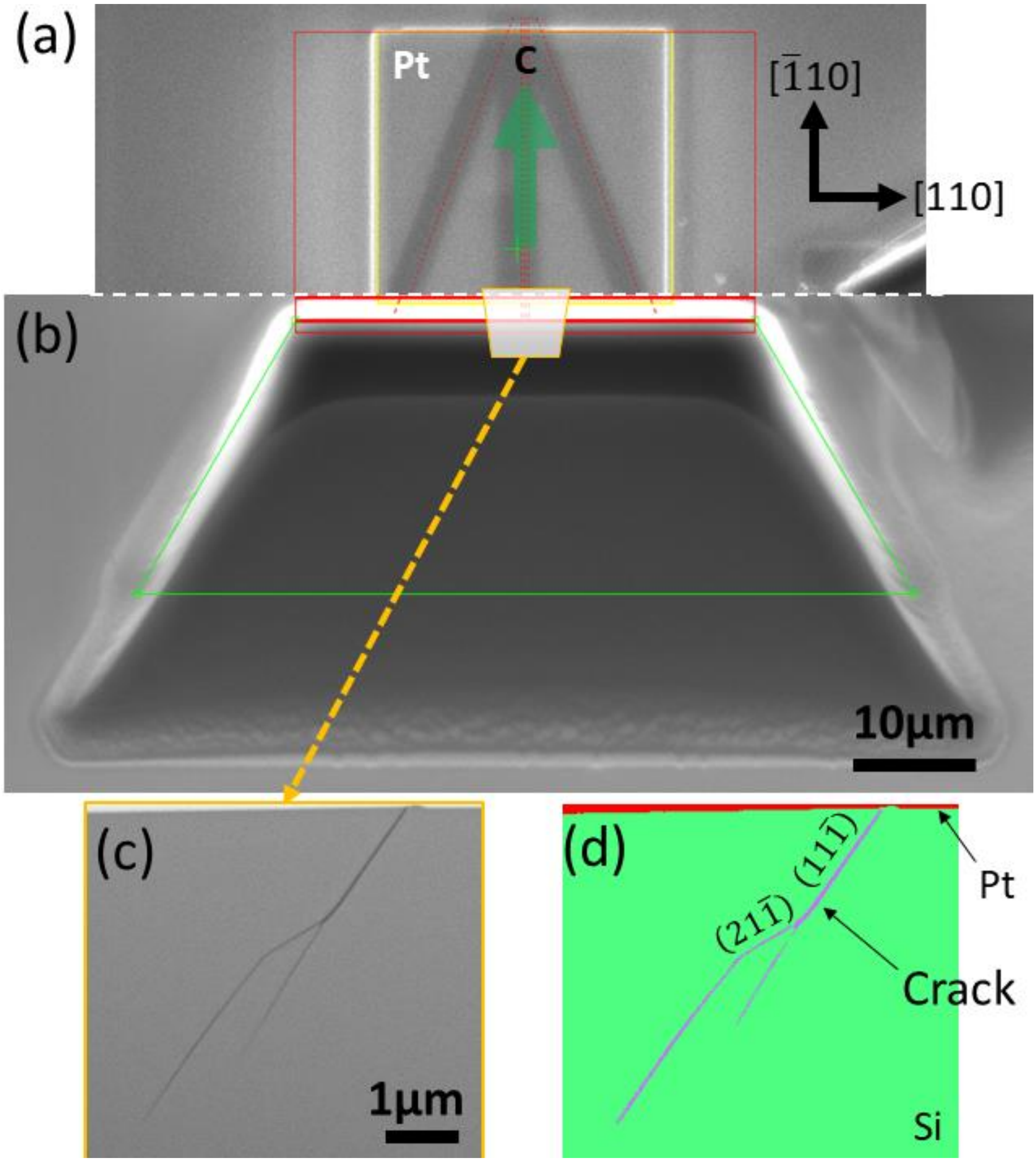

Figure 2: (a) deposition of protective Platinum (Pt) and Carbon (C) layers. (b) Trapezium trench. (c) SE image for the crack. (d) Segmented crack geometry (purple) in a Silicon (green) covered with a Pt layer (red).

The obtained 16-bit image stacks from the FIB slicing each had a voxel size of 10 x 10 x 25 $nm^3$ and examined a volume of 28.29 x 12.09 x 23.35 $\mu m^3$. 3D image drift was corrected using Fiji ImageJ [50] before manually training a Weka Segmentation classifier [51] on 20 frames to detect cracks from the matrix and the protective Pt layer (Figure 2d) before applying the trained classifier to the entire image stack. The segmented stack was then processed using AVIZO (version 2020.3.1) to visualise the crack in 3D.

# 3. Results and discussion

One crack emanated from each of the four corners of the Vickers micro-indentation. The cracks were slightly curved near the contact impression but propagated in a straight line along the [110] and $[\bar{1}10]$ directions (x and y-axis, respectively) (Figure 4a). Starting from the crack in the $[\bar{1}10]$ direction (labelled 1 in Figure 4a) and going anticlockwise until the crack labelled 4 in Figure 4a; the surface crack length ($l$) is 4.25 ± 0.05 µm, 6.70 ± 0.13 µm, 7.44 ± 0.17 µm, 5.02 ± 0.05 µm, respectively. The indentation contact's impression radius, from the centre to the corner, ($a$) is 4.42 ± 0.09 µm. All dimensions were measured from SEM images using ImageJ [50].

The fracture toughness ($K_{IC}$) was estimated for each crack using equation (1), with Young's modulus ($E$) of 165.6 GPa [46], a fitting factor of ($x_v$) of 0.0164 ± 0.004 [2] for monocrystal (001) Silicon, maximum indenter load ($P$) (obtained by multiplying the load in gf by standard gravity $g_0$), and the hardness ($H$) in MPa was approximated as $H \approx 0.4636\ {}^{Fg_0}/_{a^2}$ [52] where $F$ is the load in Kgf and $a$ in mm; this yields a hardness of 11.66 ± 0.03 GPa that agrees with reported experimental data [21,53]. The ratio ${}^{c}/_{a}$ need to be 2.5 or greater to fit the Lawn-Evans-Marshall (LEM) model [8]; this was not satisfied for cracks 1 and 4, and the estimated fracture toughness ($K_{IC}$) for crack (2) is 0.82 ± 0.14 MPa $m^{0.5}$ and crack (3) is 0.74 ± 0.12 MPa $m^{0.5}$. The variance is mainly due to uncertainties in $x_v$. These values of $K_{IC}$ are within the expected silicon fracture toughness, which varies from 0.62 to 1.29 MPa $m^{0.5}$, with {111} being the weakest plane [54,55].

The 3D observation of the cracks caused by the indentation shows they did not have the same length, straight surface trace, and did not conform to a specific cleavage system, but – in general – may they be described as having a Half-penny geometry for the $\langle 110 \rangle \{111\}$ cracks labelled (1) and (3) inclined by ~35° from the (110) near the surface and branching into a lateral crack (Figure 4b) ~3 µm away from the indentation site and deep into the sample. The half-penny crack starts from the expected location of the plastic zone under the indentation impression. The radial geometry for the $\langle 110\rangle\{110\}$ crack labelled (2) and the $\langle 530\rangle\{110\}$ crack (4) also branched to a lateral crack which can cause chipping. Radial vertical cracks started outside the plastic zone and are shallower than the half-penny cracks. The half-penny geometry of the crack switches between crystal planes, as seen in Figure 2d, where the crack changes the $(11\bar{1})$ plane to $(21\bar{1})$ plane before going back to the $(11\bar{1})$. The sporadicity of the crack-changing planes increases near the indentation plastic deformation site. Median cracks, parallel to the loading axis and induced due to the outward stress, were observed directly beneath the indentation impression and reached a maximum depth of ~1.1 µm.

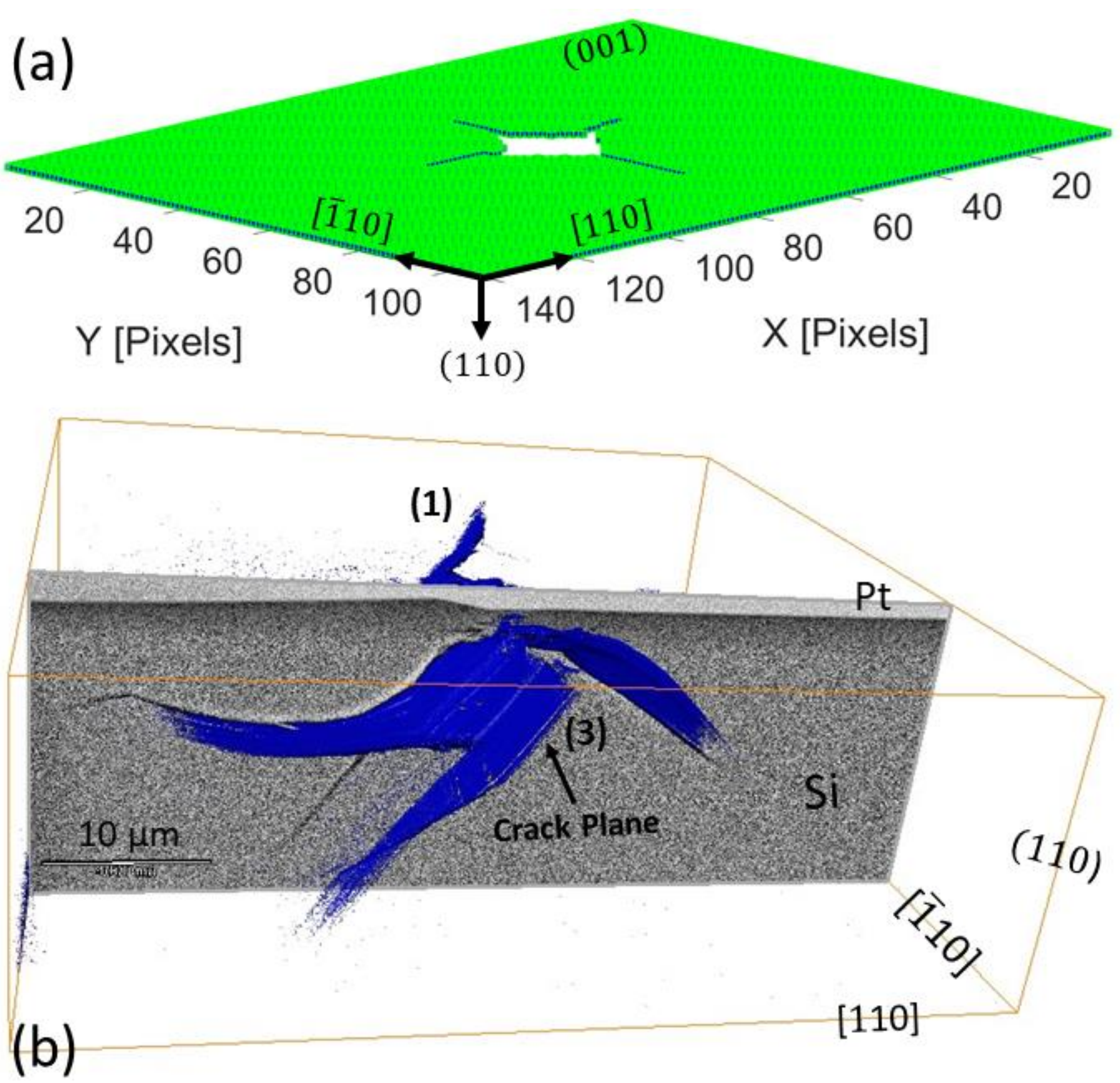

Figure 3: (a) Membrane layer, representing the probed field excluding the crack stem, meshed using eight nodes Hexahedron element. (b) Crack geometry as revealed by focused ion beam (see supplementary data).

The sequence of crack formation with indentation generally starts with median cracks, and once the indent is lifted, leaving a residual impression, the residual tensile stresses cause lateral cracks that may curve upward to meet the surface and cause chipping, whereas the residual hoop stress causes shallow radial cracks [56]. Microcracks generated by stresses from the plastic deformation at the contact impression vicinity coalesce to form a large crack that propagates towards the surface and may influence the radial crack geometry [57]. Crack deflection may also be related to the crack propagation velocity [58,59]. These factors all contribute to the irregular crack shape [21].

The elastic deformation fields (Figure 3a) were calculated by first choosing a remote reference pattern from the deformation site. The high-quality EBSPs yielded a field with an excellent average cross-correlation peak height of 0.87 ± 0.08, and an extremely low mean angular error of ± 2.82 x $10^{-4}$ rad (excluding the cracks and indentation itself). In addition, the deformation fields are symmetrical around the indentation impression, e.g., compressive normal strains can be seen along the x and y-axis with in-plane shear positive and negative along the cracks, all indicating a residual crack opening stress field.

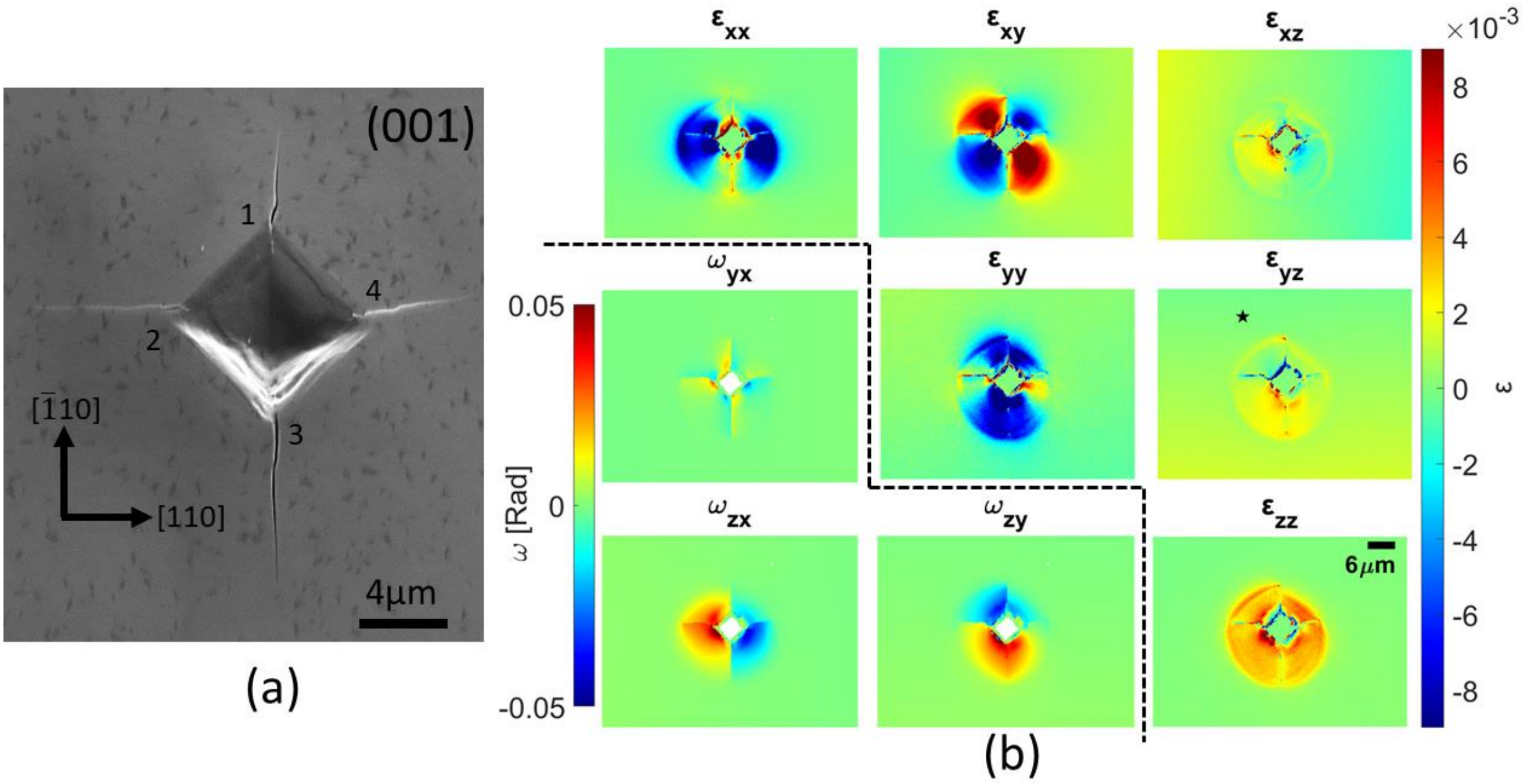

Figure 4: (a) Secondary electron microscopy (SEM) image for the indentation on the (001) mono-Si crystal. (b) HR-EBSD deviatoric strain and rotation components. The location of the reference $EBSP_0$ is highlighted with a star in $\varepsilon_{yz}$.

The out-of-plane shear strain is minimal, but there is a slight gradient in $\varepsilon_{yz}$ and $\varepsilon_{xz}$ (Figure 4b) integration to the displacement field describes a rigid body rotation. This gradient is due to the uncorrected pattern centre (PC) shift from beam movement during pattern acquisition [60]. The effect of this gradient on the integrated displacements can be removed by assigning the absolute minimum displacement to zero and then taking this as the origin to extract the rotation angles $(\psi, \theta, \phi)$ using Kevin Shoemaker's method [63]. This allows one to construct the transformation matrix ($R$ in equation 13) and correct the displacement field's rigid body movement [64]. Figure 5 shows the inverse relationship between the induced rigid body rotation and thickness (Z) of the HR-EBSD information depth.

$$R = \begin{bmatrix} \cos\theta\cos\phi & \cos\theta\sin\phi & -\sin\theta \\ \sin\psi\sin\theta\cos\phi - \cos\psi\sin\phi & \sin\psi\sin\theta\sin\phi + \cos\psi\cos\theta & \cos\theta\sin\psi \\ \cos\psi\sin\theta\cos\phi + \sin\psi\sin\phi & \cos\psi\sin\theta\sin\phi - \sin\psi\cos\phi & \cos\theta\cos\psi \end{bmatrix} \qquad 13$$

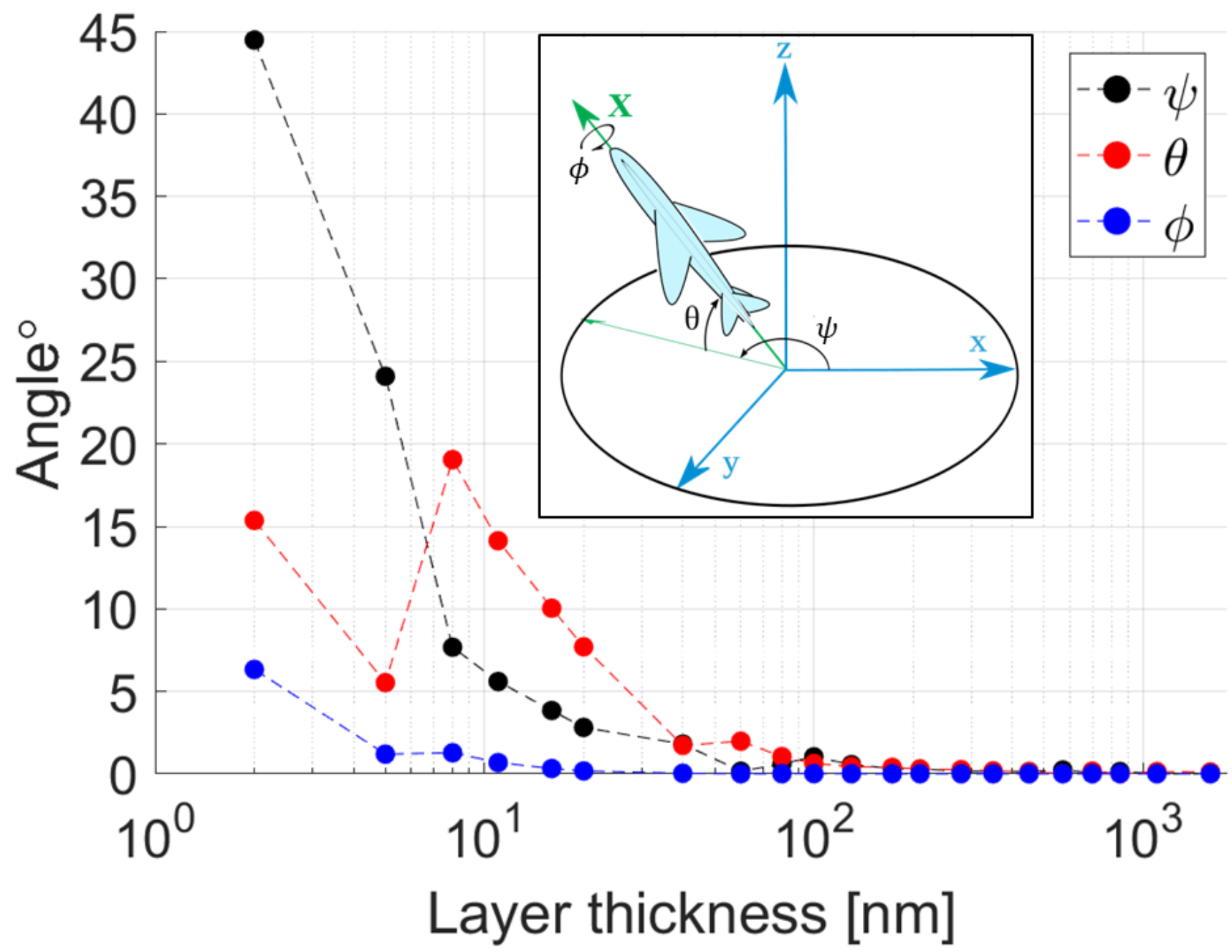

Figure 5: Corrected rigid body Euler angles $(\psi, \theta, \phi)$ extracted from the out of plane shear strain field gradients of Figure 4b, .as a function of the assumed layer thickness, Z, of HR-EBSD information

The out-of-plane (positive) normal strain distribution ($\varepsilon_{zz}$ in Figure 4b) is similar to out-of-plane surface displacements that have been obtained from optical interference measurements of a Palmqvist crack system [11] and correlate with the calculated GND density distribution (Figure 6a). The GND density near the indentation impression edges is 7.7 ± 1.8 x $10^{12}$, decreasing gradually to about 2.7 ± 0.9 x$10^{12}$ $m^{-2}$ with 83% of the dislocations at 45°/$\langle 110 \rangle b$ to the surface normal (between 1 and 2 in Figure 6) before falling to a background density of 0.3 ± 0.1 x $10^{12}$ $m^{-2}$. Silicon is a brittle material at room temperature, but when indented, the deformation is accompanied by a competitive process of transformation-induced plasticity [13,14] and dislocation nucleation at the surface [61,62]. As the indenter moves into the sample, uniform material pile-up occurs around the indentation, changing the local orientation and strain status compared to undeformed material away from the indenter contact site. The GND represents lattice rotations from the distortion of the crystal and does not necessarily indicate actual plasticity. Thus, the calculated geometrically necessary dislocations (Figure 7a) represent equivalent dislocations that maintain the material change in orientation. Interestingly, the GND distribution (Figure 6a) is elliptical and rotated by 1.2 from the edge trace of the indentation ($\theta$ in Figure 6a). This may be due to the indenter being slightly inclined relative to the surface, which would affect the measured field and cracks lengths.

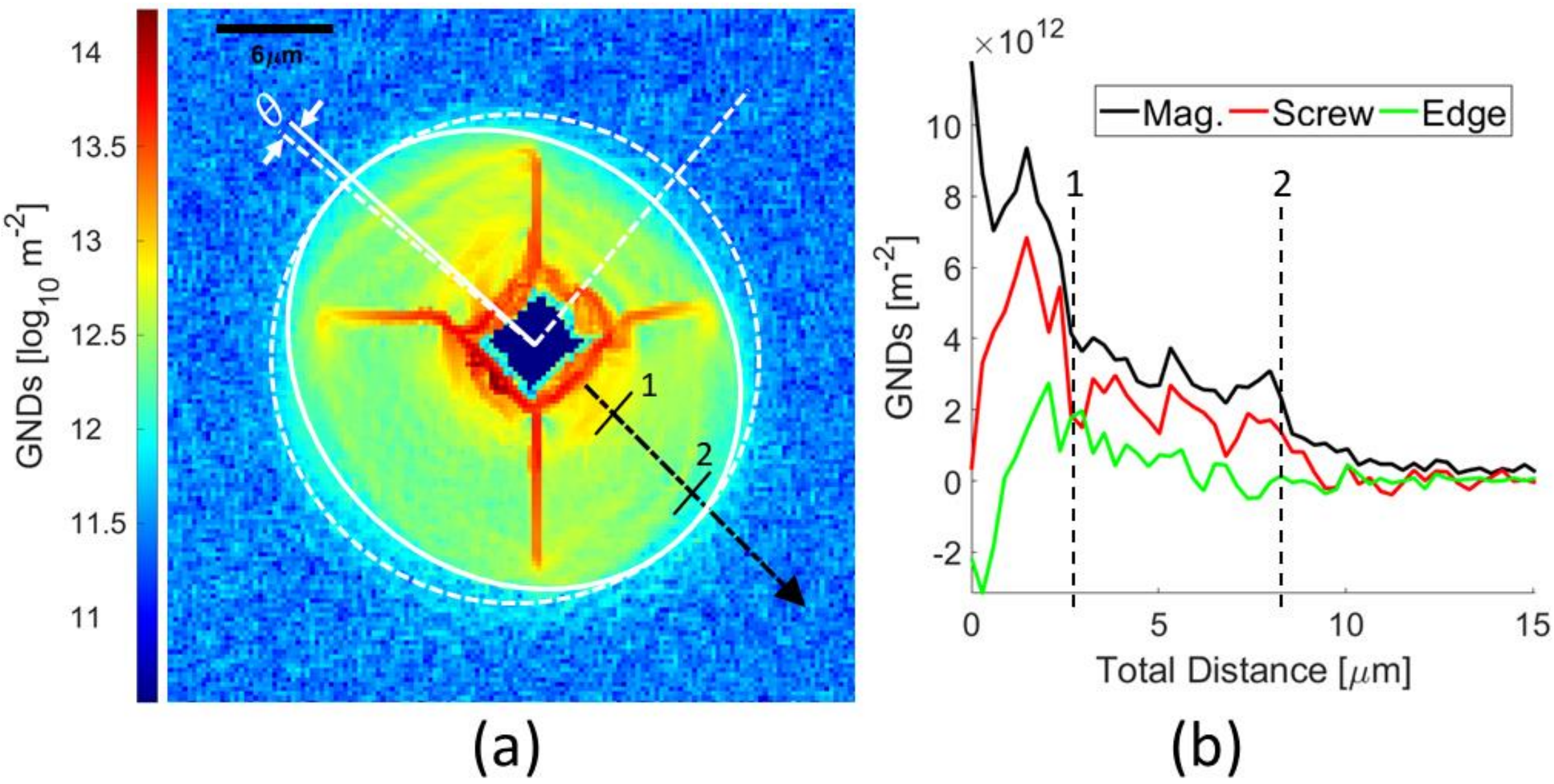

Figure 6: (a) Estimated geometrically necessary dislocations (GND) density with the theoretical circular profile around the indentation in dashed white lines and experiment with an oblique profile in continuous white lines. The angle between the circle and oblique is $\theta$. Dashed black line is where the GND density line profile shown in (b) is taken, starting from near the indentation impression and going outward.

The elastic strain fields were integrated into the equivalent displacement fields, assuming different ranges of the EBSP information depth ($Z$). The profile of the out-of-plane displacement $U_z$ increases towards the indenter centre, similar to the GND density profile discussed earlier. The magnitude of the $U_z$ changed with the assumed membrane thickness, as did the spatial distribution (Figure 7). However, the in-plane displacements ($U_x$ and $U_y$) were identical in magnitude and did not change with Z. Results for Z=700 nm are shown in Figure 8b to d. The in-plane displacements (Figure 8c and d) indicate that the cracks may be under a residual tensile opening; it is generally assumed that once the indenter was removed, the cracks should be minimally loaded [4,40,41] only by the residual tensile stresses exerted due to the compressed plastic zone. This assumption will be investigated shortly.

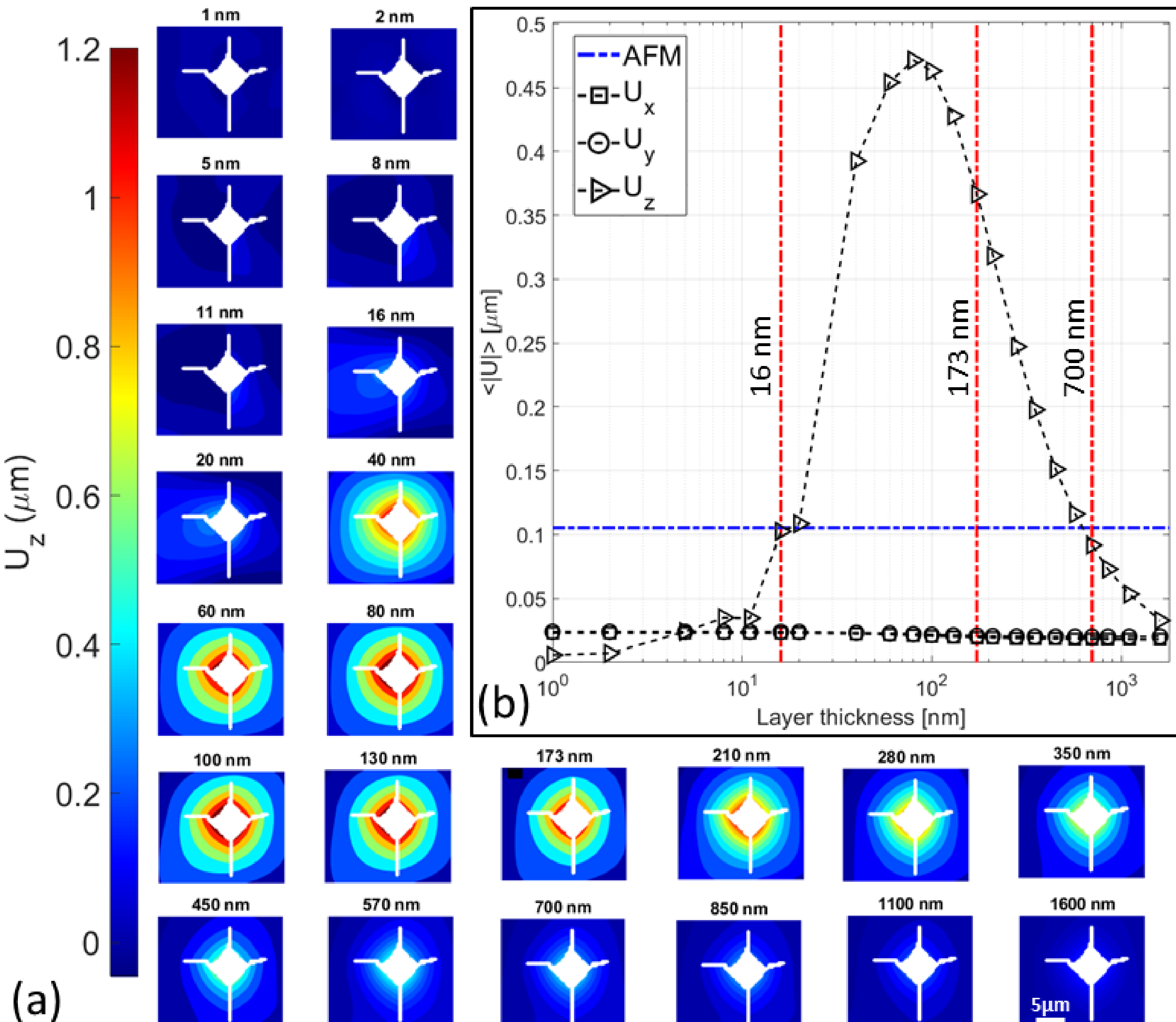

Figure 7: Displacement Integration assuming different membrane (Z) thickness illustrated as (a) $U_z$ maps (b) absolute average. The integrated displacement fields' absolute average was calculated from a window of 25 x 25 μm$^2$ around the indent to match the AFM window.

The out-of-plane displacement fields, calculated using a range of assumed depth Z, were compared to the topographical profile of the indentation impression measured using AFM. Agreement in magnitude was found at $Z$ of 700 nm and 16 nm (Figure 7b), but the distributions agreed best at 700 nm, which is also the depth within which 91.7% of the signal is produced. The actual depth resolution cannot be concluded as deep as 700 nm just by virtue of the Uz profile (obtained as the integrated elastic displacement from the elastic strains) being matched with the AFM profile. The depth resolution of EBSD is widely accepted to vary between 10 to 40 nm, decreasing with the material atomic number [63]. AFM measures the indentation impression profile, which is also affected by plasticity [64]. Hence the significant value of the apparent depth resolution may be caused by these additional displacements. However, some experimental studies [65] have concluded that depth resolution could extend to 1000 nm due to inelastic scattering, so the apparent depth of 700 nm may be reasonable.

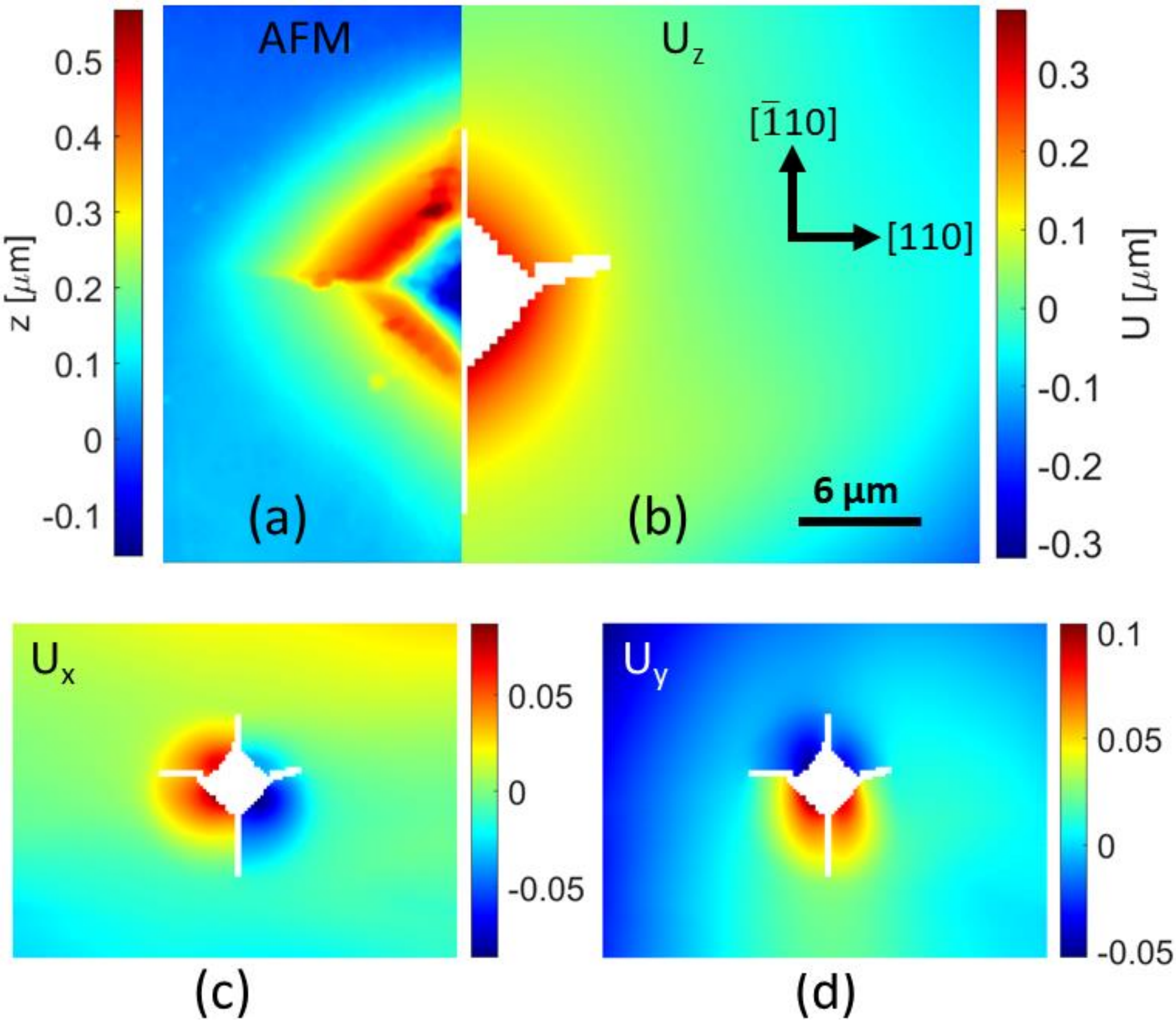

Figure 8: (a) AFM measured topography around the indentation impression. Integrated (b) $U_z$, (c) $U_x$ and (d) $U_y$ elastic displacement calculated while assuming 700 nm depth (Z).

The depth resolution of EBSD is widely accepted to vary between 10 to 40 nm, decreasing with the material atomic number [63]. Nevertheless, experimental measurement, using a differently thick transparent amorphous layer of Cr coating a mono-Si crystal, indicated that the depth of resolution is as shallow as 2 nm, determined by Si pattern quality deteriorating by ~50% when using a FEG-SEM with 15 kV beam conditions and 15 mm working distance between the beam and sample and 65 mm between the sample and the detector and without considering the channelling effect [66]. Using a similar experimental approach, different results were reported (Table 1), e.g., Isabell and David [65] concluded that depth resolution could extend to 1 µm due to inelastic scattering (including tangential smearing and channelling effect).

Table 1: Measured EBSP depth resolution from experiments and Monte Carlo (MC) based simulation to infer EBSP depth resolution.

| Author | Material | Density ($kg/m^3$)[3] | Voltage (kV) | Depth Resolution (nm) |
|---|---|---|---|---|
| Dingley [63] | – | – | – | 10-40 |
| Dingley and Randle [67] | – | – | – | 10 |
| El-Dasher *et al.* [68] | – | – | – | ~20 |
| Bhattacharyya and Eades [69] | – | – | – | < 1000 (MC) |
| Zaefferer [66] | Si | 2.33 | 15 | 3.5 ± 1.5 |
| Yamamoto [70] | Al | 2.70 | 20 | 50 |
| Bhattacharyya and Eades [69] | Al | 2.70 | 20 | 15-40 |
| Baba-Kishi [71] | Al | 2.70 | 20 | > 50 |
| Ren *et al.* [72] | Al | 2.70 | 20 | 115 (MC) |
| Isabell and Dravid [65] | Al | 2.70 | 30 | ~400 |
| Michael and Goehner [73,74] | Al | 2.70 | 40 | 100 |
| Yamamoto [70] | Al | 2.70 | 50 | 120 |
| Isabell and Dravid [65] | Nb | 3.58 | 30 | < 1000 |
| Isabell and Dravid [65] | $SrTiO_3$ | 5.11 | 30 | ~300 |
| Keller *et al.* [75] | GaAs | 5.32 | 15 | 30 |
| Steinmetz and Zaefferer [76] | Fe | 7.87 | 7.5 | 10 |

[3] At room temperature.

| Author | Material | Density (kg/m$^3$) | Voltage (kV) | Depth Resolution (nm) |
|---|---|---|---|---|
| Bordín *et al.* [77] | σ-Fe | 7.87 | 20 | 16 (MC) |
| C. Zhu *et al.* [78] | Ni | 8.91 | 10 | 10 (MC) |
| Harland *et al.* [79][4] | Ni | 8.91 | 30 | ≲ 10 |
| Kohl [65,80] | Ni | 8.91 | 30 | 5-6 (MC) |
| Michael and Goehner [73,74] | Ni | 8.91 | 40 | 20 |
| Chen et al. [81] | Cu | 8.96 | 5 | 38 |
| Chen *et al.* [81] | Cu | 8.96 | 10 | 46 |
| Yamamoto [70] | Cu | 8.96 | 20 | 20 |
| Ren *et al.* [72] | Cu | 8.96 | 20 | 35 (MC) |
| Chen *et al.* [81] | Cu | 8.96 | 30 | 72 |
| Yamamoto [70] | Cu | 8.96 | 50 | 50 |
| Ren *et al.* [72] | Ag | 10.49 | 20 | 30 (MC) |
| Isabell and Dravid [65] | W | 19.30 | 30 | ~50 |
| Ren *et al.* [72] | Au | 19.30 | 20 | 22 (MC) |
| Harland *et al.* [82] | Au | 19.30 | 30 | 80 |
| Michael and Goehner [73,74] | Au | 19.30 | 40 | 10 |

MC simulation results seem more consistent, decreasing with the material density, as the calculated depth of resolution for EBSPs formation is understood using Block wave theory where backscattered primary electrons, after interacting with the crystal lattice, exit the surface carrying information about the crystallinity of volume that is interacting with the electrons. The backscattered electrons (BSE) energy distribution depends on the material's characteristics and the beam conditions [83]. This BSE wave field is also affected by the thermal diffuse scattering process that causes incoherent and inelastic (energy loss) scattering – after the Bragg diffraction events – which does not, yet, have a complete physical description that can be related to mechanisms that constitute EBSP depth resolution [84,85]. However, there are supporting arguments that MC simulation gives an (accurate [80,86]) approximation that is based on the wrong assumptions [66].

[4] Using small angle detector.

On the contrary, the experimental results in Table 1 are not consistent. These experiments are highly cumbersome due to the need for highly precise and well-calibrated equipment, with the results open to interpretation [87]. This is mainly because there is no agreement about the definition/criteria of depth resolution. For example, definitions that are dependent on where ~92% of the signal is generated [88,89], pattern quality [66], or as ambiguous as "*where useful information is obtained*" [90]. All reported values in Table 1 either do not mention a definition or do not have a rationale for the definition. In addition, most of these experiments do not mention the beam size, tilt angle, beam-to-sample and sample-to-detector working distance, and – sometimes – even the beam energy, which are critical parameters for determining (or simulating) the depth resolution of the patterns as the interaction volume increases with beam energy and size and decreases with the sample atomic number or density [65]. Also, the beam current is mostly not considered a parameter that can affect the depth resolution, neither in experimental nor in simulation. However, it affects the beam spot size and pattern signal-to-noise (S/N) ratio [78,91,92]. Most importantly, conclusions drawn from both experiments and simulations assumed the surface is pristine and heterogeneity of the depth resolution, which are not valid for a deformed sample [66].

Determining the depth resolution is still challenging, as it was made clear from the contradicting depth resolution reported in the literature; thus, we will assume the depth resolution as 700 nm (where 91.7% of the information is coming, Figure 1) considering that this depth resolution yields (elastic) out-of-plane displacement that is similar to AFM.

To characterise the indentation cracks, we use the elastic strain field obtained from HR-EBSD to calculate the stress intensity factors for the inclined $(11\bar{1})$ crack labelled (3) and the orthogonal $(\bar{1}10)$ crack labelled (2). The three-dimensional stress intensity factors (SIFs) can be extracted using the interaction integral natively implemented in ABAQUS® finite element solve, which uses the displacement field obtained from integrating the elastic strains at 700 nm depth of information as a boundary condition and the material's elastic stress-strain relationship [36][5]. Plane stress conditions were assumed by considering the thin probed layer and that the sample is not constrained at the surface; thus, it deforms freely in the third dimension [93]. The interaction integral approach – as implemented in ABAQUS® – is robust [94] and less sensitive to the crack position compared to field fitting approaches [95].

The field at crack 3 (the $(11\bar{1})$ crack) was mainly compressive $\varepsilon_{xx}$, with a minimal $\varepsilon_{yy}$ (localised at the crack tip) and a uniform in-plane shear strain along the crack. The strains $\varepsilon_{xx}$ and $\varepsilon_{zz}$ uniformly encapsulated the crack (Figure 9a). The analysis of the field, when calculated for integration contours where convergence stabilised (shaded area in Figure 9b), obtained a total mode I stress intensity factor ($K_I^T$) of 0.02 ± 0.00 MPa $m^{0.5}$, including the in-plane component of mode I ($K_I$) that equals 0.01 ± 0.00 MPa $m^{0.5}$, in-plane mode II shear ($K_{II}$) of -0.38 ± 0.00 MPa $m^{0.5}$, and out-of-plane mode III shear ($K_{III}$) of -0.38 ± 0.01 MPa $m^{0.5}$. The (negative) sign of the in-plane $K_{II}$ and out-of-plane $K_{III}$ shear depends on the arrangement of the nodes at the tip and does not carry any physical meaning [42]. Therefore, the crack has a residual shear loading, in a direction inclined to the indented surface, with a negligible opening load.

[5] Code and example are available at https://doi.org/10.5281/zenodo.6411568.

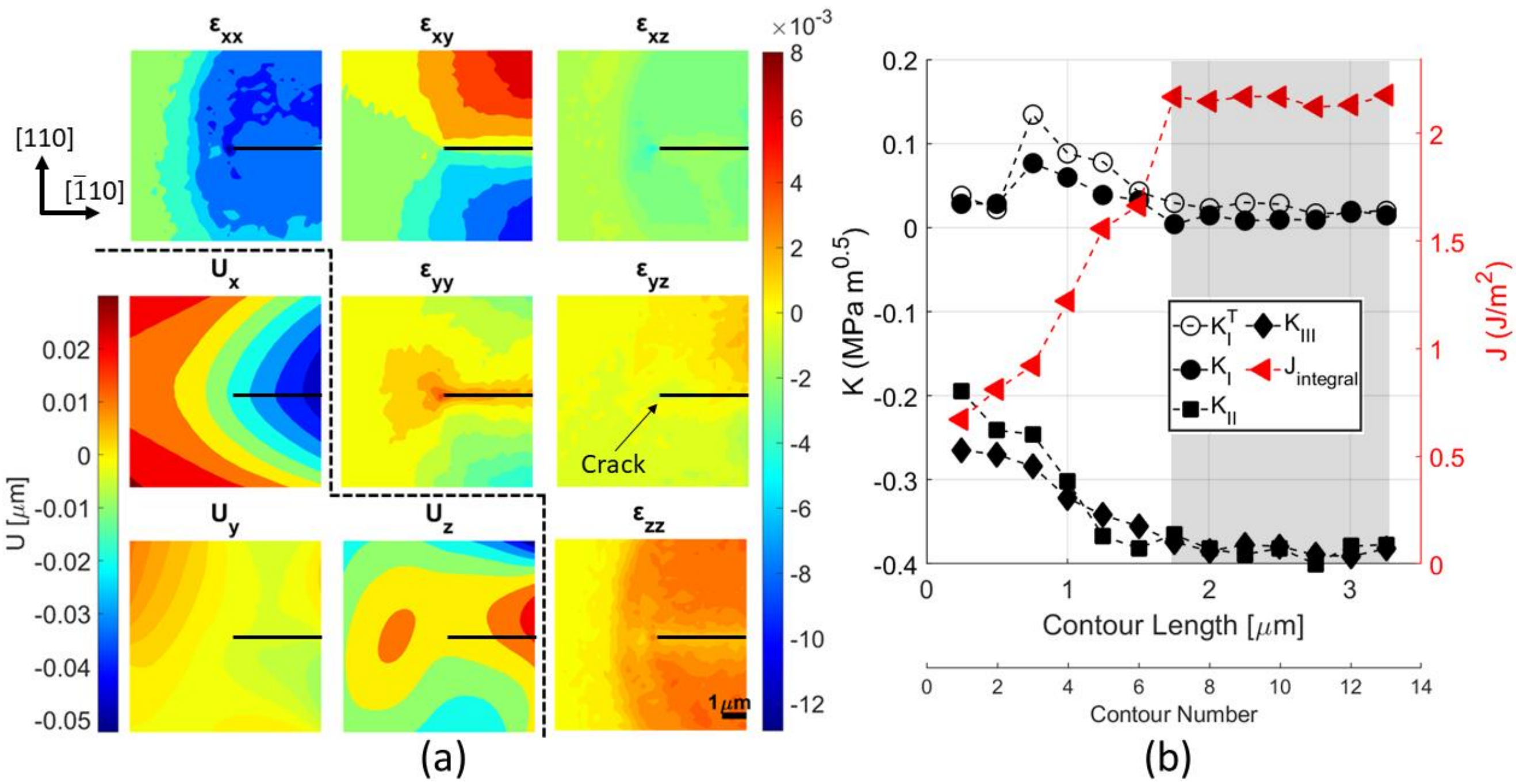

Figure 9: (a) Elastic strain and displacement fields for $(11\bar{1})$ crack number 3 in Figure 4a. Displacement was integrated, assuming a 700 nm depth resolution. (b) *J*-integral and stress intensity factors were calculated from the crack field.

On the other hand, crack 2 (the $(\bar{1}10)$ crack) has a field with similar in-plane and out-of-plane shear strain but without an apparent strain $\varepsilon_{yy}$ localisation at the crack tip. The strains $\varepsilon_{xx}$ and $\varepsilon_{zz}$ are also not uniformly distributed around the crack. The crack plane is orthogonal to the surface and is loaded by a $K_I^T$ of -0.31 ± 0.04 MPa $m^{0.5}$, $K_I$ of -0.19 ± 0.03 MPa $m^{0.5}$, $K_{II}$ of 0.26 ± 0.02 MPa $m^{0.5}$, and $K_{III}$ of -0.66 ± 0.01 MPa $m^{0.5}$.

The higher out-of-plane mode III ($K_{III}$) and the out-of-plane component of mode I ($K_I^r$) in these examples, inflate the value of the strain energy release rate (*J*-integral) to extend it can be higher than those measured for loaded cracks [96]. This is because $K_{III}$ and $K_I^r$ are dependent on the depth resolution and the size of the integrated window [97]. Thus, these values should be dismissed unless correction methods are developed to encounter these dependencies.

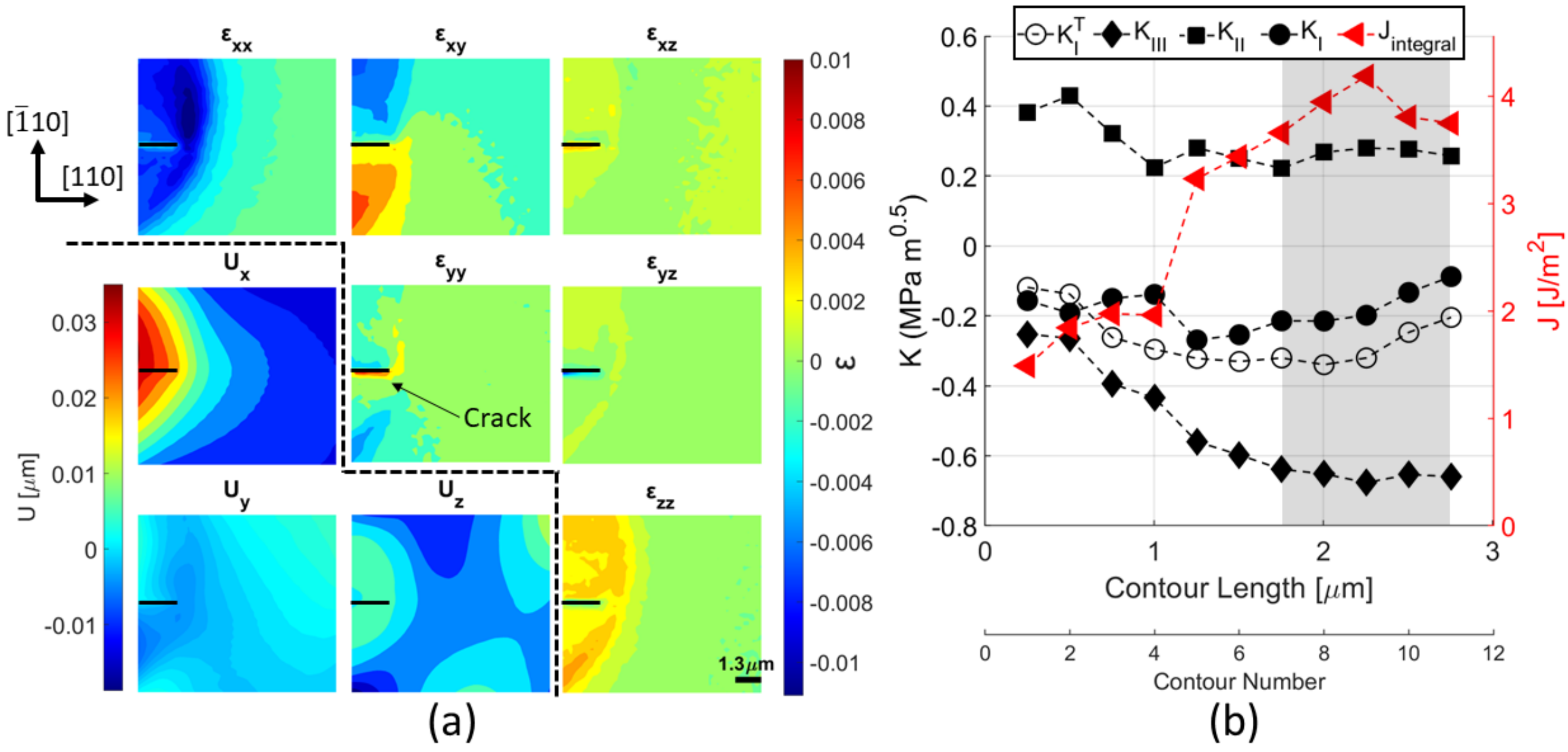

Figure 10: (a) Elastic strain and displacement fields for $(\bar{1}10)$ crack number 2 in Figure 4a. Displacement was integrated, assuming a 700 nm depth resolution. (b) *J*-integral and stress intensity factors were calculated from the crack field.

Although it is somewhat speculative to draw conclusions from these analyses; the observation of mode I in crack (2), which is orthogonal to the surface, and the higher mode II of crack (3), which is inclined to the surface, is interesting. This difference in the mechanical conditions ahead of these cracks may be due to the orientation of the crack plane geometry relative to the indentation. Both cracks have the same directional Young's modulus and Poisson's ratio [46] due to similar forces from the indentation. The difference in crack planes between crack (2) and (3) is speculated to be due to the relationship between the plane normalised stress to the plane fracture toughness, i.e., for crack (2) $\sigma_{\perp\{110\}}/K_{Ic\{110\}} > \sigma_{\perp\{111\}}/K_{Ic\{111\}}$ [98] and surface cracks mixed modality may be due to stress relaxation from lateral cracking [99].

Considering the crack shape, the measured surface mechanical conditions will change with depth; an in situ three-dimensional strain map will be optimal to fully characterise the crack field and properly link it to geometry, especially in materials where the local fracture behaviour needs further investigation. In principle, this might be achieved by a 3D strain measurement technique such as Laue micro-diffraction [100] performed in situ, or it might be done using in situ EBSD with a sample geometry that allows examination of the crack under

load – such as indentation close to an edge (see [101]). Thus, the method discussed here can be used with (three-dimensional [99,100,102]) local deformation measurement to calculate fracture toughness without detailed knowledge of the indentation deformation process or the crack length; thus, analysing short cracks that do not fit Lawn-Evans-Marshall (LEM) model.

# 4. Conclusion

A novel approach was derived to compute the elastic displacement field from a measured elastic deformation field (i.e., deformation gradient or strain). The method is based on integrating the deformation field using finite element discretisation, and was applied to investigate the deformation and cracks of Vickers micro-indentation on a (001) mono-Si crystal sample.

The elastic deformation fields at the indented surface, calculated from the measured electron backscattered patterns (EBSPs), reveal a symmetrical deformation field around the indentation impression. The elastic strain was integrated into the equivalent displacements field; with the assumption of a depth resolution of 700 nm, the out-of-plane displacement field agreed with the topography measured by atomic force microscopy (AFM).

The stress intensity factors (SIFs) were calculated from the local deformation field at the crack vicinity. This indicated that an opening mode I loading existed at the crack orthogonal to the surface, and a significant in-plane shear mode II existed, with no mode I, for the inclined crack.

## Acknowledgements

We thank Professor Peter Wilshaw (University of Oxford) for supplying the specimen material and Alexander J. Leide (University of Bristol) for help with nano-indentation. The authors acknowledge the use of experimental equipment from the Oxford Materials Characterisation Service (OMCS) and characterisation facilities within the David Cockayne Centre for Electron Microscopy (DCCEM), Department of Materials, University of Oxford, alongside financial support provided by the Henry Royce Institute (Grant ref EP/R010145/1). Abdalrhaman Koko thanks Engineering and Physical Sciences Research Council (EPSRC) for providing PhD studentship (Grant ref EP/N509711/1).

## Authorship Contribution Statement

**Abdalrhaman Koko:** Conceptualization, Methodology, Software, Investigation, Formal analysis, Writing - original draft, Visualisation.

**Elsiddig Elmukashfi:** Methodology, Software, Writing - original draft.

**Phani S. Karamched:** Investigation, Visualisation.

**Angus J. Wilkinson:** Software.

**Thomas James Marrow:** Resources, Writing - review & editing, Supervision, Funding Acquisition.

## References

[1] Lawn B, Wilshaw R. Indentation fracture: principles and applications. J Mater Sci 1975;10:1049–81. https://doi.org/10.1007/BF00823224.

[2] Anstis GR, Chantikul P, Lawn BR, Marshall DB. A Critical Evaluation of Indentation Techniques for Measuring Fracture Toughness: I, Direct Crack Measurements. Journal of the American Ceramic Society 1981;64:533–8. https://doi.org/10.1111/j.1151-2916.1981.tb10320.x.

[3] Marshall DB, Lawn BR, Evans AG. Elastic/Plastic Indentation Damage in Ceramics: The Lateral Crack System. Journal of the American Ceramic Society 1982;65:561–6. https://doi.org/10.1111/J.1151-2916.1982.TB10782.X.

[4] Cook RF, Gerbig Y, Schoenmaker J, Stranick S. Stress-intensity factor and toughness measurement at the nanoscale using confocal Raman microscopy. 12th International Conference on Fracture 2009, ICF-12 2009;5:3758–66.

[5] Lawn B, Wilshaw R. Indentation fracture: principles and applications. Journal of Materials Science 1975 10:6 1975;10:1049–81. https://doi.org/10.1007/BF00823224.

[6] Evans AG, Charles EA. Fracture Toughness Determinations by Indentation. Journal of the American Ceramic Society 1976;59:371–2. https://doi.org/10.1111/J.1151-2916.1976.TB10991.X.

[7] Marshall DB, Lawn BR. Residual stress effects in sharp contact cracking. Journal of Materials Science 1979 14:8 1979;14:2001–12. https://doi.org/10.1007/BF00551043.

[8] Lawn BR, Evans AG, Marshall DB. Elastic/Plastic Indentation Damage in Ceramics: The Median/Radial Crack System. Journal of the American Ceramic Society 1980;63:574–81. https://doi.org/10.1111/j.1151-2916.1980.tb10768.x.

[9] Cuadrado N, Casellas D, Anglada M, Jiménez-Piqué E. Evaluation of fracture toughness of small volumes by means of cube-corner nanoindentation. Scr Mater 2012;66:670–3. https://doi.org/https://doi.org/10.1016/j.scriptamat.2012.01.033.

[10] Dukino RD, Swain M v. Comparative Measurement of Indentation Fracture Toughness with Berkovich and Vickers Indenters. Journal of the American Ceramic Society 1992;75:3299–304. https://doi.org/10.1111/j.1151-2916.1992.tb04425.x.

[11] Laugier MT. Palmqvist indentation toughness in WC-Co composites. Journal of Materials Science Letters 1987 6:8 1987;6:897–900. https://doi.org/10.1007/BF01729862.

[12] Quinn GD, Bradt RC. On the Vickers Indentation Fracture Toughness Test. Journal of the American Ceramic Society 2007;90:673–80. https://doi.org/10.1111/J.1551-2916.2006.01482.X.

[13] Jiapeng S, Cheng L, Han J, Ma A, Fang L. Nanoindentation Induced Deformation and Pop-in Events in a Silicon Crystal: Molecular Dynamics Simulation and Experiment. Sci Rep 2017;7:10282. https://doi.org/10.1038/s41598-017-11130-2.

[14] Goel S, Haque Faisal N, Luo X, Yan J, Agrawal A. Nanoindentation of polysilicon and single crystal silicon: Molecular dynamics simulation and experimental validation. J Phys D Appl Phys 2014;47:275304. https://doi.org/10.1088/0022-3727/47/27/275304.

[15] Morrell BREBLL; R. Measurement Good Practice Guide No 9: Palmqvist Toughness for Hard and Brittle Materials. Teddington: 2008.

[16] Liu M, Lin JY, Lu C, Tieu KA, Zhou K, Koseki T. Progress in Indentation Study of Materials via Both Experimental and Numerical Methods. Crystals 2017, Vol 7, Page 258 2017;7:258. https://doi.org/10.3390/CRYST7100258.

[17] Tanaka M, Higashida K, Nakashima H, Takagi H, Fujiwara M. Orientation dependence of fracture toughness measured by indentation methods and its relation to surface energy in single crystal silicon. International Journal of Fracture 2006 139:3 2006;139:383–94. https://doi.org/10.1007/S10704-006-0021-7.

[18] Qin J, Huang Y, Hwang KC, Song J, Pharr GM. The effect of indenter angle on the microindentation hardness. Acta Mater 2007;55:6127–32. https://doi.org/10.1016/J.ACTAMAT.2007.07.016.

[19] Britton TB, Liang H, Dunne FPE, Wilkinson AJ. The effect of crystal orientation on the indentation response of commercially pure titanium: experiments and simulations. Proceedings of the Royal Society A: Mathematical, Physical and Engineering Sciences 2010;466:695–719. https://doi.org/10.1098/RSPA.2009.0455.

[20] Ebrahimi F, Kalwani L. Fracture anisotropy in silicon single crystal. Materials Science and Engineering: A 1999;268:116–26. https://doi.org/10.1016/S0921-5093(99)00077-5.

[21] Cook RF. Strength and sharp contact fracture of silicon. J Mater Sci 2006;41:841–72. https://doi.org/10.1007/s10853-006-6567-y.

[22] Lapitskaya VA, Kuznetsova TA, Khabarava A v., Chizhik SA, Aizikovich SM, Sadyrin E v., et al. The use of AFM in assessing the crack resistance of silicon wafers of various orientations. Eng Fract Mech 2022;259:107926. https://doi.org/10.1016/J.ENGFRACMECH.2021.107926.

[23] Zhu P, Zhao Y, Agarwal S, Henry J, Zinkle SJ. Toward accurate evaluation of bulk hardness from nanoindentation testing at low indent depths. Mater Des 2022;213:110317. https://doi.org/10.1016/J.MATDES.2021.110317.

[24] Li Z, Ghosh A, Kobayashi AS, Bradt RC. Indentation Fracture Toughness of Sintered Silicon Carbide in the Palmqvist Crack Regime. Journal of the American Ceramic Society 1989;72:904–11. https://doi.org/10.1111/j.1151-2916.1989.tb06242.x.

[25] Pharr GM, Herbert EG, Gao Y. The Indentation Size Effect: A Critical Examination of Experimental Observations and Mechanistic Interpretations. Annu Rev Mater Res 2010;40:271–92. https://doi.org/10.1146/annurev-matsci-070909-104456.

[26] Broitman E. Indentation Hardness Measurements at Macro-, Micro-, and Nanoscale: A Critical Overview. Tribol Lett 2017;65:1–18. https://doi.org/10.1007/S11249-016-0805-5/FIGURES/16.

[27] Barhli SM, Saucedo-Mora L, Simpson C, Becker T, Mostafavi M, Withers PJ, et al. Obtaining the J-integral by diffraction-based crack-field strain mapping. Procedia Structural Integrity 2016;2:2519–26. https://doi.org/10.1016/j.prostr.2016.06.315.

[28] Huber JE, Hofmann F, Barhli S, Marrow TJ, Hildersley C. Observation of crack growth in a polycrystalline ferroelectric by synchrotron X-ray diffraction. Scr Mater 2017;140:23–6. https://doi.org/10.1016/j.scriptamat.2017.06.053.

[29] Barhli SM, Saucedo-Mora L, Jordan MSL, Cinar AF, Reinhard C, Mostafavi M, et al. Synchrotron X-ray characterisation of crack strain fields in polygranular graphite. Carbon N Y 2017;124:357–71. https://doi.org/10.1016/J.CARBON.2017.08.075.

[30] Koko A, Earp P, Wigger T, Tong J, Marrow TJ. J-integral analysis: An EDXD and DIC comparative study for a fatigue crack. Int J Fatigue 2020;134:105474. https://doi.org/10.1016/j.ijfatigue.2020.105474.

[31] Friedman LH, Vaudin MD, Stranick SJ, Stan G, Gerbig YB, Osborn W, et al. Assessing strain mapping by electron backscatter diffraction and confocal Raman microscopy using wedge-indented Si. Ultramicroscopy 2016;163:75–86. https://doi.org/10.1016/J.ULTRAMIC.2016.02.001.

[32] Cherepanov GP. Crack propagation in continuous media. Journal of Applied Mathematics and Mechanics 1967;31:503–12. https://doi.org/10.1016/0021-8928(67)90034-2.

[33] Rice JR. A Path Independent Integral and the Approximate Analysis of Strain Concentration by Notches and Cracks. J Appl Mech 1968;35:379–86. https://doi.org/10.1115/1.3601206.

[34] Britton TB, Wilkinson AJ. Stress fields and geometrically necessary dislocation density distributions near the head of a blocked slip band. Acta Mater 2012;60:5773–82. https://doi.org/10.1016/j.actamat.2012.07.004.

[35] Guo Y, Britton TB, Wilkinson AJ. Slip band–grain boundary interactions in commercial-purity titanium. Acta Mater 2014;76:1–12. https://doi.org/10.1016/J.ACTAMAT.2014.05.015.

[36] Koko A, Elmukashfi E, Becker TH, Karamched PS, Wilkinson AJ, Marrow TJ. In situ characterisation of the strain fields of intragranular slip bands in ferrite by high-

resolution electron backscatter diffraction. Acta Mater 2022;239:118284. https://doi.org/10.1016/J.ACTAMAT.2022.118284.

[37] Guo Y. The interactions between slip band, deformation twins and grain boundaries in commercial purity titanium. University of Oxford, 2015.

[38] Koko A, Elmukashfi E, Dragnevski K, Wilkinson AJ, Marrow TJ. J-integral analysis of the elastic strain fields of ferrite deformation twins using electron backscatter diffraction. Acta Mater 2021;218:117203. https://doi.org/10.1016/j.actamat.2021.117203.

[39] Koko A, Becker TH, Elmukashfi E, Pugno NM, Wilkinson AJ, Marrow TJ. HR-EBSD analysis of in situ stable crack growth at the micron scale. J Mech Phys Solids 2023;172:105173. https://doi.org/10.1016/j.jmps.2022.105173.

[40] Leide AJ, Todd RI, Armstrong DEJ. Effect of Ion Irradiation on Nanoindentation Fracture and Deformation in Silicon Carbide. JOM 2021;73:1617–28. https://doi.org/10.1007/S11837-021-04636-8/FIGURES/8.

[41] Porporati AA, Hosokawa K, Zhu W, Pezzotti G. Stress dependence of the cathodoluminescence spectrum of N-doped 3C-SiC. J Appl Phys 2006;100:093508. https://doi.org/10.1063/1.2363260.

[42] Systèmes® D. ABAQUS. ABAQUS v66 2009. https://classes.engineering.wustl.edu/2009/spring/mase5513/abaqus/docs/v6.6/index.html.

[43] Wilkinson AJ, Meaden G, Dingley DJ. High-resolution elastic strain measurement from electron backscatter diffraction patterns: New levels of sensitivity. Ultramicroscopy 2006;106:307–13. https://doi.org/10.1016/j.ultramic.2005.10.001.

[44] Britton TB, Wilkinson AJ. Measurement of residual elastic strain and lattice rotations with high resolution electron backscatter diffraction. Ultramicroscopy 2011;111:1395–404. https://doi.org/10.1016/j.ultramic.2011.05.007.

[45] Koko A, Wilkinson AJ, Marrow TJ. An iterative method for reference pattern selection in high resolution electron backscatter diffraction (HR-EBSD). Ultramicroscopy (Under Review) 2022. https://doi.org/https://doi.org/10.48550/arXiv.2206.10242.

[46] Hopcroft MA, Nix WD, Kenny TW. What is the Young's Modulus of Silicon? Journal of Microelectromechanical Systems 2010;19:229–38. https://doi.org/10.1109/JMEMS.2009.2039697.

[47] Wilkinson AJ, Randman D. Determination of elastic strain fields and geometrically necessary dislocation distributions near nanoindents using electron back scatter diffraction. Philosophical Magazine 2010;90:1159–77. https://doi.org/10.1080/14786430903304145.

[48] Hovington P, Drouin D, Gauvin R. CASINO: A new monte carlo code in C language for electron beam interaction —part I: Description of the program. Scanning 1997;19:1–14. https://doi.org/https://doi.org/10.1002/sca.4950190101.

[49] Tang YT, D'Souza N, Roebuck B, Karamched P, Panwisawas C, Collins DM. Ultra-high temperature deformation in a single crystal superalloy: Mesoscale process simulation and micromechanisms. Acta Mater 2021;203:116468. https://doi.org/10.1016/J.ACTAMAT.2020.11.010.

[50] Collins TJ. ImageJ for microscopy. Biotechniques 2007;43:S25–30. https://doi.org/10.2144/000112517.

[51] Arganda-Carreras I, Kaynig V, Rueden C, Eliceiri KW, Schindelin J, Cardona A, et al. Trainable Weka Segmentation: a machine learning tool for microscopy pixel classification. Bioinformatics 2017;33:2424–6. https://doi.org/10.1093/bioinformatics/btx180.

[52] ISO 6507-1:2005 - Metallic materials — Vickers hardness test — Part 1: Test method. 2005.

[53] Ebrahimi F, Kalwani L. Fracture anisotropy in silicon single crystal. Materials Science and Engineering: A 1999;268:116–26. https://doi.org/10.1016/S0921-5093(99)00077-5.

[54] Masolin A, Bouchard P-O, Martini R, Bernacki M. Thermo-mechanical and fracture properties in single-crystal silicon. J Mater Sci 2013;48:979–88. https://doi.org/10.1007/s10853-012-6713-7.

[55] Pérez R, Gumbsch P. Directional Anisotropy in the Cleavage Fracture of Silicon. Phys Rev Lett 2000;84:5347–50. https://doi.org/10.1103/PhysRevLett.84.5347.

[56] Fischer-Cripps AC. Other Techniques in Nanoindentation. In: Ling FF, editor. Nanoindentation. 3rd ed., New York, NY: Springer New York; 2011, p. 163–80. https://doi.org/10.1007/978-1-4419-9872-9_9.

[57] Connally JA, Brown SB. Slow Crack Growth in Single-Crystal Silicon. Science (1979) 1992;256:1537–9. https://doi.org/10.1126/science.256.5063.1537.

[58] Sherman D, Be'ery I. Velocity dependent crack deflection in single crystal silicon. Scr Mater 2003;49:551–5. https://doi.org/https://doi.org/10.1016/S1359-6462(03)00354-3.

[59] Sherman D, Be'ery I. From crack deflection to lattice vibrations—macro to atomistic examination of dynamic cleavage fracture. J Mech Phys Solids 2004;52:1743–61. https://doi.org/https://doi.org/10.1016/j.jmps.2004.02.004.

[60] Britton TB, Jiang J, Guo Y, Vilalta-Clemente A, Wallis D, Hansen LN, et al. Tutorial: Crystal orientations and EBSD — Or which way is up? Mater Charact 2016;117:113–26. https://doi.org/10.1016/j.matchar.2016.04.008.

[61] Minor AM, Lilleodden ET, Jin M, Stach EA, Chrzan DC, Morris JW. Room temperature dislocation plasticity in silicon. Https://DoiOrg/101080/14786430412331315680 2011;85:323–30. https://doi.org/10.1080/14786430412331315680.

[62] Gouldstone A, Chollacoop N, Dao M, Li J, Minor AM, Shen YL. Indentation across size scales and disciplines: Recent developments in experimentation and modeling. Acta Mater 2007;55:4015–39. https://doi.org/10.1016/J.ACTAMAT.2006.08.044.

[63] Dingley D. Progressive steps in the development of electron backscatter diffraction and orientation imaging microscopy. J Microsc 2004;213:214–24. https://doi.org/10.1111/J.0022-2720.2004.01321.X.

[64] Danilewsky A, Wittge J, Kiefl K, Allen D, McNally P, Garagorri J, et al. Crack propagation and fracture in silicon wafers under thermal stress. J Appl Crystallogr 2013;46:849–55. https://doi.org/10.1107/S0021889813003695/XZ5004SUP1.WMV.

[65] Isabell TC, Dravid VP. Resolution and sensitivity of electron backscattered diffraction in a cold field emission gun SEM. Ultramicroscopy 1997;67:59–68. https://doi.org/https://doi.org/10.1016/S0304-3991(97)00003-X.

[66] Zaefferer S. On the formation mechanisms, spatial resolution and intensity of backscatter Kikuchi patterns. Ultramicroscopy 2007;107:254–66. https://doi.org/https://doi.org/10.1016/j.ultramic.2006.08.007.

[67] Dingley DJ, Randle V. Microtexture determination by electron back-scatter diffraction. Journal of Materials Science 1992 27:17 1992;27:4545–66. https://doi.org/10.1007/BF01165988.

[68] El-Dasher BS, Adams BL, Rollett AD. Viewpoint: experimental recovery of geometrically necessary dislocation density in polycrystals. Scr Mater 2003;48:141–5. https://doi.org/10.1016/S1359-6462(02)00340-8.

[69] Bhattacharyya A, Eades JA. Use of an energy filter to improve the spatial resolution of electron backscatter diffraction. Scanning 2009;31:114–21. https://doi.org/10.1002/SCA.20150.

[70] Yamamoto T. Experimental aspects of electron channeling patterns in scanning electron microscopy. II. Estimation of contrast depth. Physica Status Solidi (a) 1977;44:467–76. https://doi.org/10.1002/PSSA.2210440208.

[71] Baba-Kishi KZ. Measurement of crystal parameters on backscatter kikuchi diffraction patterns. Scanning 1998;20:117–27. https://doi.org/10.1002/SCA.1998.4950200210.

[72] Ren SX, Kenik EA, Alexander KB, Goyal A. Exploring Spatial Resolution in Electron Back-Scattered Diffraction Experiments via Monte Carlo Simulation. Microscopy and Microanalysis 1998;4:15–22. https://doi.org/10.1017/S1431927698980011.

[73] Schwarzer RA, Field DP, Adams BL, Kumar M, Schwartz AJ. Present State of Electron Backscatter Diffraction and Prospective Developments. In: Schwartz AJ, Kumar M, Adams BL, Field DP, editors. Electron Backscatter Diffraction in Materials Science, vol. 53, Boston, MA: Springer US; 2009, p. 1–20. https://doi.org/10.1007/978-0-387-88136-2_1.

[74] Engler O, Randle V. Introduction to Texture Analysis. 2nd ed. CRC Press; 2009. https://doi.org/10.1201/9781420063660.

[75] Keller RR, Roshko A, Geiss RH, Bertness KA, Quinn TP. EBSD measurement of strains in GaAs due to oxidation of buried AlGaAs layers. Microelectron Eng 2004;75:96–102. https://doi.org/https://doi.org/10.1016/j.mee.2003.11.010.

[76] Steinmetz DR, Zaefferer S. Towards ultrahigh resolution EBSD by low accelerating voltage. Https://DoiOrg/101179/026708309X12506933873828 2013;26:640–5. https://doi.org/10.1179/026708309X12506933873828.

[77] Bordín SF, Limandri S, Ranalli JM, Castellano G. EBSD spatial resolution for detecting sigma phase in steels. Ultramicroscopy 2016;171:177–85. https://doi.org/10.1016/J.ULTRAMIC.2016.09.010.

[78] Zhu C, de Graef M. EBSD pattern simulations for an interaction volume containing lattice defects. Ultramicroscopy 2020;218:113088. https://doi.org/10.1016/J.ULTRAMIC.2020.113088.

[79] Harland CJ, Akhter P, Venables JA. Accurate microcrystallography at high spatial resolution using electron back-scattering patterns in a field emission gun scanning electron microscope. J Phys E 1981;14:175. https://doi.org/10.1088/0022-3735/14/2/011.

[80] Dingley DJ, Field DP. Electron backscatter diffraction and orientation imaging microscopy. Materials Science and Technology 1997;13:69–78. https://doi.org/10.1179/mst.1997.13.1.69.

[81] Chen D, Kuo J-C, Wu W-T. Effect of microscopic parameters on EBSD spatial resolution. Ultramicroscopy 2011;111:1488–94. https://doi.org/10.1016/j.ultramic.2011.06.007.

[82] Harland CJ, Klein JH, Akhter P, Venables JA. Electron back-scattering patterns in a field emission gun scanning electron microscope. Proceedings of the 9th International Congress on Electron Microscopy, Toronto 1981;1:564–5.

[83] Brodusch N, Demers H, Gauvin R. Imaging with a Commercial Electron Backscatter Diffraction (EBSD) Camera in a Scanning Electron Microscope: A Review. J Imaging 2018;4. https://doi.org/10.3390/jimaging4070088.

[84] Tanaka M, Terauchi M, Tsuda K, Saitoh K. Convergent-beam electron diffraction IV. vol. 2. Jeol; 2002.

[85] Winkelmann A. Dynamical Simulation of Electron Backscatter Diffraction Patterns BT - Electron Backscatter Diffraction in Materials Science. In: Schwartz AJ, Kumar M, Adams BL, Field DP, editors., Boston, MA: Springer US; 2009, p. 21–33. https://doi.org/10.1007/978-0-387-88136-2_2.

[86] Ren SX, Kenik EA, Alexander KB, Goyal A. Exploring Spatial Resolution in Electron Back-Scattered Diffraction Experiments via Monte Carlo Simulation. Microscopy and Microanalysis 1998;4:15–22. https://doi.org/10.1017/S1431927698980011.

[87] Wisniewski W, Rüssel C. An experimental viewpoint on the information depth of EBSD. Scanning 2016;38:164–71. https://doi.org/https://doi.org/10.1002/sca.21251.

[88] Powell CJ, Jablonski A. Surface Sensitivity of Auger-Electron Spectroscopy and X-ray Photoelectron Spectroscopy. Journal of Surface Analysis 2011;17:170–6. https://doi.org/10.1384/JSA.17.170.

[89] PiŇos J, MikmekovÁ, Frank L. About the information depth of backscattered electron imaging. J Microsc 2017;266:335–42. https://doi.org/10.1111/JMI.12542.

[90] ISO 18115:2001/Amd 2:2007 - Surface chemical analysis — Vocabulary — Amendment 2. Geneva: 2007.

[91] Humphreys FJ. Characterisation of fine-scale microstructures by electron backscatter diffraction (EBSD). Scr Mater 2004;51:771–6. https://doi.org/10.1016/J.SCRIPTAMAT.2004.05.016.

[92] Goldstein JI, Newbury DE, Michael JR, Ritchie NWM, Scott JHJ, Joy DC. The Visibility of Features in SEM Images. Scanning Electron Microscopy and X-Ray Microanalysis, New York, NY: Springer New York; 2018, p. 123–31. https://doi.org/10.1007/978-1-4939-6676-9_8.

[93] Hardin TJ, Ruggles TJ, Koch DP, Niezgoda SR, Fullwood DT, Homer ER. Analysis of traction-free assumption in high-resolution EBSD measurements. J Microsc 2015;260:73–85. https://doi.org/https://doi.org/10.1111/jmi.12268.

[94] Courtin S, Gardin C, Bézine G, ben Hadj Hamouda H. Advantages of the J-integral approach for calculating stress intensity factors when using the commercial finite element software ABAQUS. Eng Fract Mech 2005;72:2174–85. https://doi.org/10.1016/j.engfracmech.2005.02.003.

[95] Becker TH, Mostafavi M, Tait RB, Marrow TJ. An approach to calculate the J-integral by digital image correlation displacement field measurement. Fatigue Fract Eng Mater Struct 2012;35:971–84. https://doi.org/10.1111/j.1460-2695.2012.01685.x.

[96] Masolin A, Bouchard P-O, Martini R, Bernacki M. Thermo-mechanical and fracture properties in single-crystal silicon. J Mater Sci 2013;48:979–88. https://doi.org/10.1007/s10853-012-6713-7.

[97] Koko A, Elmukashfi E, Becker TH, Karamched PS, Wilkinsona AJ, Marrow TJ. Full-field high-resolution EBSD strain analysis of intragranular slip bands in ferrite. Acta Materialia (Under Review) 2022.

[98] Moulins A, Ma L, Dugnani R, Zednik RJ. Dynamic crack modeling and analytical stress field analysis in single-crystal silicon using quantitative fractography. Theoretical and Applied Fracture Mechanics 2020;109:102693. https://doi.org/10.1016/J.TAFMEC.2020.102693.

[99] Vertyagina Y, Mostafavi M, Reinhard C, Atwood R, Marrow TJ. In situ quantitative three-dimensional characterisation of sub-indentation cracking in polycrystalline alumina. J Eur Ceram Soc 2014;34:3127–32. https://doi.org/10.1016/J.JEURCERAMSOC.2014.04.002.

[100] MacDowell AA, Celestre RS, Tamura N, Spolenak R, Valek B, Brown WL, et al. Submicron X-ray diffraction. Nucl Instrum Methods Phys Res A 2001;467–468:936–43. https://doi.org/https://doi.org/10.1016/S0168-9002(01)00530-7.

[101] Laurie Palasse; Jaroslav Lukeš. In-situ SEM nanoindentation combined with 3D EBSD. 2021.

[102] Kutsal M, Bernard P, Berruyer G, Cook PK, Hino R, Jakobsen AC, et al. The ESRF dark-field x-ray microscope at ID06. IOP Conf Ser Mater Sci Eng 2019;580:12007. https://doi.org/10.1088/1757-899x/580/1/012007.